\documentclass[review, fleqn, 12pt]{elsarticle}

 \usepackage{graphics}
\usepackage{graphicx}
\usepackage{epsfig}
\usepackage{subfig}
\usepackage{amssymb}
\usepackage{amsthm}
\usepackage{amsmath}

\newtheorem{theorem}{Theorem}[section]

\journal{Elsevier}

\begin{document}

\begin{frontmatter}



\title{Solitary-wave solutions to a dual equation of the Kaup-Boussinesq system}


\author[]{Jiangbo Zhou\corref{cor1}}
\ead{zhoujiangbo@yahoo.cn}

\author[]{Lixin Tian}
\author[]{Xinghua Fan}

\cortext[cor1]{Corresponding author. Tel: +86-511-88969336, Fax:
+86-511-88969336 }
\address{Nonlinear Scientific Research Center, Faculty of Science, Jiangsu University,
 Zhenjiang,  Jiangsu 212013,  China}
\begin{abstract}
In this paper, we employ the bifurcation theory of planar dynamical
systems to investigate the travelling-wave solutions to a dual
equation of the Kaup-Boussinesq system. The expressions for smooth
solitary-wave solutions are obtained.
\end{abstract}

\begin{keyword} dual equation of the Kaup-Boussinesq system \sep  solitary-wave solution\sep
bifurcation method


\MSC 35Q51 \sep 34C23 \sep 37G10 \sep 35Q35
\end{keyword}
\end{frontmatter}


\section{Introduction}
\label{} \setcounter{equation}{0}

Since the theory of solitons has very wide applications in fluid
dynamics, nonlinear optics, biochemistry, microbiology, physics and
many other fields, the study of soliton solutions to nonlinear
partial differential equations has become an increasingly important
area of research \cite{1}-\cite{5}. It is known that, solitons are
the solitary waves that retain their individuality under interaction
and eventually travel with their original shapes and speeds.
Therefore, to investigate the soliton solutions, one must firstly
obtain the solitary-wave solutions. Many efforts have been denoted
to seeking solitary-wave solutions to nonlinear partial differential
equations (see, e.g., \cite{6}-\cite{13}).

Recently, Guha \cite{14} studied the dual counter part of the
following Kaup-Boussinesq system \cite{15},
\begin{equation}
\label{eq1.1} \left\{ {\begin{array}{l}
u_t=v_{xxx}+2(uv)_x, \\
 v_t=u_x+2vv_x, \\
 \end{array}} \right.
\end{equation}
where $u(x, t)$ denotes the height of the water surface above a
horizontal bottom and $v(x, t)$ is related to the horizontal
velocity field. Using moment of inertia operators method and the
frozen Lie-Poisson structure, Guha derived the dual counter part of
system (\ref{eq1.1}), that is
\begin{equation}
\label{eq1.2} \left\{ {\begin{array}{l}
m_t+(mu+\frac{1}{2}u_x^2+u^2+2uv)_x=0, \\
p_t+(pu)_x=0, \\
 \end{array}} \right.
\end{equation}
where $m=u_{xx}+\mu u+\lambda v$, $p=\lambda u+v$. System
(\ref{eq1.2}) is a two component integrable system \cite{14}. When
$\mu=\lambda=0$, it becomes
\begin{equation}
\label{eq1.3} \left\{ {\begin{array}{l}
u_{xxt}+(uu_{xx}+\frac{1}{2}u_x^2+u^2+2uv)_x=0, \\
v_t+(uv)_x=0. \\
 \end{array}} \right.
\end{equation}

Various aspects of the Kaup-Boussinesq system (\ref{eq1.1}) have
been studied. For instance, Smirnov \cite{16} obtained real finite
gap regular solutions to system (\ref{eq1.1}), and Borisov et al.
\cite{17} applied the proliferation scheme to system (\ref{eq1.1}).
Also the closely related variant of system (\ref{eq1.1}) have been
studied intensively (see \cite{18}-\cite{29}). However, it seems
that the dual equation of system (\ref{eq1.1}) has attracted little
attention.

In this paper, we use the bifurcation theory of planar dynamical
systems (see \cite{30}-\cite{33}) to investigate the travelling-wave
solutions to system (\ref{eq1.3}) and obtain analytic expressions
for its smooth solitary-wave solutions. To the best of our
knowledge, the solitary-wave solutions to system (\ref{eq1.3}) have
not been reported in the literature. The bifurcation method was
first used by Li and Liu \cite{34} to obtain smooth and non-smooth
travelling-wave solutions to a nonlinearly dispersive equation and
was later employed by many authors to derive a variety of
travelling-wave solutions to a large number of nonlinear partial
differential equations \cite{35}-\cite{42}.

The remainder of the paper is organized as follows. In Section 2,
using the travelling-wave ansatz, we transform system (\ref{eq1.3})
into a planar dynamical system and then discuss bifurcations of
phase portraits of this system. In Section 3, we obtain the
expressions for smooth solitary-wave solutions to system
(\ref{eq1.3}). A short conclusion is given in Section 4.

\section{Bifurcation analysis}
 \label{}
 \setcounter{equation}{0}

Let $\xi=x-ct$, where $c$ is the wave speed. By using the
travelling-wave ansatz $u(x, t) =\varphi (x-ct)= \varphi (\xi )$,
$v(x, t) =\psi (x-ct)= \psi (\xi )$, we reduce system (\ref{eq1.3})
to the following ordinary differential equations:
\begin{equation}
\label{eq2.1} \left\{ {\begin{array}{l}
- c\varphi''' +(\varphi \varphi''+\frac{1}{2}\varphi'^2 +\varphi^2+2 \varphi \psi)'=0,\\
 -c\psi'+(\varphi\psi)'=0.\\
 \end{array}} \right.
\end{equation}
Integrating (\ref{eq2.1}) once with respect to $\xi$, we have
\begin{equation}
\label{eq2.2} \left\{ {\begin{array}{l}
- c\varphi'' +\varphi \varphi''+\frac{1}{2}\varphi'^2 +\varphi^2+2 \varphi \psi=g_1,\\
 -c\psi+\varphi\psi=g_2, \\
 \end{array}} \right.
\end{equation}
where $g_1, g_2$ are two integral constants.

From the second equation in system (\ref{eq2.2}), we obtain
\begin{equation}
\label{eq2.3} \psi=\frac{g_2}{\varphi-c}.
\end{equation}
Substituting (\ref{eq2.3}) into the first equation in system
(\ref{eq2.2}) yields
\begin{equation}
\label{eq2.4}  \varphi'' =\frac{(
\varphi-c)(g_1-\frac{1}{2}\varphi'^2- \varphi^2)-2 g_2
\varphi}{(\varphi - c)^2}.
\end{equation}
Let $ \varphi '= \frac {\sqrt{2}}{2}y $, then we get the following
planar dynamical system:
\begin{equation}
\label{eq2.5}\left\{ {\begin{array}{l}
 \frac{\textstyle  d\varphi }{\textstyle  d\xi } = \frac{\sqrt{2}}{2} y, \\
 \frac{\textstyle dy}{\textstyle d\xi } =\frac{\textstyle \sqrt{2}((
\varphi-c)(g_1-\frac{1}{2}y^2- \varphi^2)-2 g_2
\varphi)}{\textstyle(\varphi - c)^2}. \\
 \end{array}} \right.
\end{equation}
\noindent This is a planar Hamiltonian system with Hamiltonian
function
\begin{equation}
\label{eq2.6} H(\varphi,
y)=\varphi^4-\frac{4c}{3}\varphi^3+(4g_2-2g_1)\varphi^2+4cg_1\varphi+(\varphi-c)^2y^2=h,
\end{equation}
where $h$ is a constant.

Note that (\ref{eq2.5}) has a singular line $\varphi = c$. To avoid
the line temporarily we make transformation $d\xi =
\frac{\sqrt{2}}{2}(\varphi - c)^2 d\zeta $. Under this
transformation, Eq.(\ref{eq2.5}) becomes
\begin{equation}
\label{eq2.7} \left\{ {\begin{array}{l}
 \frac{\textstyle d\varphi }{\textstyle d\zeta } = \frac{1}{2}(\varphi - c)^2y, \\
 \frac{\textstyle dy}{\textstyle d\zeta } =(\varphi-c)(g_1-\frac{1}{2}y^2- \varphi^2)-2 g_2
\varphi.
\\
 \end{array}} \right.
\end{equation}

System (\ref{eq2.5}) and system (\ref{eq2.7}) have the same first
integral as (\ref{eq2.6}). Consequently, system (\ref{eq2.7}) has
the same topological phase portraits as system (\ref{eq2.5}) except
for the straight line $\varphi = c$.

For a fixed $h$, (\ref{eq2.6}) determines a set of invariant curves
of system (\ref{eq2.7}). As $h$ is varied, (\ref{eq2.6}) determines
different families of orbits of system (\ref{eq2.7}) having
different dynamical behaviors. Let $M(\varphi _e , y_e )$ be the
coefficient matrix of the linearized version of system (\ref{eq2.7})
at the equilibrium point $(\varphi _e , y_e )$, then
\begin{equation}
\label{eq2.8} M(\varphi _e ,y_e ) = \left( {{\begin{array}{*{20}c}
{\indent  \indent (\varphi _e-c) y_e } \hfill & {\indent  \frac{1}{2}(\varphi _e - c)^2} \hfill \\
 {-3\varphi _e^2 +2c \varphi _e+g_1-2g_2-\frac{1}{2}y^2 } \hfill & \indent{-(\varphi _e-c)y_e} \hfill \\
\end{array} }} \right)
\end{equation}
and at this equilibrium point, we have
\begin{equation}
\begin{split}
\label{eq2.9} J(\varphi _e ,y_e ) &= \det M(\varphi _e ,y_e )
\\& = - (\varphi _e-c)^2 y_e^2+\frac{1}{2}(\varphi _e -
c)^2 (3\varphi _e^2 -2c \varphi _e-g_1+2g_2+\frac{1}{2}y^2),
\end{split}
\end{equation}
\begin{equation}
\label{eq2.10} p(\varphi _e ,y_e ) = \mathrm{trace}(M(\varphi _e
,y_e )) = 0.
\end{equation}

It is easy to see that the equilibrium point of system (\ref{eq2.7})
is in the form of $(\varphi_e, 0)$. At this equilibrium point, we
have $J(\varphi _e , 0)= \frac{1}{2}(\varphi _e - c)^2 (3\varphi
_e^2 -2c \varphi _e-g_1+2g_2) $. By using the bifurcation theory of
planar dynamical system, we know that if $J(\varphi_e, 0)>0 $ (or $
< 0$), then the equilibrium $(\varphi_e, 0)$ is a center (or saddle)
point; if $J(\varphi_e, 0) =0$, and  the Poincar\'{e} index of the
equilibrium point is 0, then it is a cusp.

Usually, a solitary-wave solution to system (\ref{eq1.3})
corresponds to a homoclinic orbit of system (\ref{eq2.7}).
Therefore, to obtain solitary-wave solutions to system
(\ref{eq1.3}), we need only to seek homoclinic orbits of system
(\ref{eq2.7}) and so only the saddle points are of interest.
Firstly, we need to look for the possible zeros of the function
\begin{equation}
\label{eq2.8} f(\varphi)= -\varphi^3+c \varphi^2+a\varphi+b,
\end{equation}
where $a=g_1-2g_2$, $b=-cg_1$.

In order to find all possible zeros of $f(\varphi)$, we set
\begin{equation}
\label{eq2.9} f'(\varphi)= -3\varphi^2+2c \varphi+a=0.
\end{equation}
When $\Delta=4c^2+12a>0$, we find two real roots to Eq.(\ref{eq2.9})
as follows:
\begin{equation}
\label{eq2.10} \varphi_1^*=\frac{1}{3}(c- \sqrt{c^2+3a}),
\end{equation}
\begin{equation}
\label{eq2.11} \varphi_2^*=\frac{1}{3}(c+ \sqrt{c^2+3a}),
\end{equation}
with $\varphi_1^*<\varphi_2^*$. When $\Delta=4c^2+12a\leq 0$, the
inequality  $f'(\varphi)\leq 0$ holds. In this case, if $(\varphi_e,
0)$ is an equilibrium point of system (\ref{eq2.7}), then it is a
center point (or a cusp) because $J(\varphi _e , 0) =
-\frac{1}{2}(\varphi _e - c)^2 f'(\varphi_e)\geq 0$. Therefore, in
the following, we always suppose $\Delta=4c^2+12a> 0$.

 Substitute (\ref{eq2.10}) and (\ref{eq2.11}) into
(\ref{eq2.8}), respectively, we get
\begin{equation}
\label{eq2.12}
f_1=f(\varphi_1^*)=\frac{2c^3}{27}-\frac{2c^2}{27}\sqrt{c^2+3a}-\frac{2a}{9}\sqrt{c^2+3a}+\frac{ac}{3}+b,
\end{equation}
\begin{equation}
\label{eq2.13}
f_2=f(\varphi_2^*)=\frac{2c^3}{27}+\frac{2c^2}{27}\sqrt{c^2+3a}+\frac{2a}{9}\sqrt{c^2+3a}+\frac{ac}{3}+b,
\end{equation}
with
\begin{equation}
\label{eq2.12} f_1-f_2=-\frac{2c^2}{27}\sqrt{c^2+3a}(c^2+3a)<0.
\end{equation}

The equilibrium points of system (\ref{eq2.7}) have the following
properties.

\begin{theorem}
\label{th1} (1) If $f_2 < 0$, then system (\ref{eq2.7}) has only one
equilibrium point, denoted by $(\varphi_1, 0)
(\varphi_1<\varphi_1^*<\varphi_2^*)$, which is a center point;

(2) If $f_1 > 0$, then system (\ref{eq2.7}) has only one equilibrium
point, denoted by $(\varphi_2, 0)
(\varphi_1^*<\varphi_2^*<\varphi_2)$, which is a center point;

(3) If  $f_1 = 0$, then system (\ref{eq2.7}) has two equilibrium
points, denoted by $(\varphi_3, 0), (\varphi_4, 0)
(\varphi_3=\varphi_1^*<\varphi_2^*<\varphi_4)$. $(\varphi_3, 0)$ is
a cusp, while $(\varphi_4, 0)$ is a center point;

(4) If  $f_2= 0$, then system (\ref{eq2.7}) has two equilibrium
points, denoted by $(\varphi_5, 0), (\varphi_6, 0)
(\varphi_5<\varphi_1^*<\varphi_2^*=\varphi_6)$. $(\varphi_5, 0)$ is
a center point, while $(\varphi_6, 0)$ is a cusp;

(5) If  $f_1 < 0, f_2>0$, then system (\ref{eq2.7}) has three
equilibrium points, denoted by $(\varphi_7, 0), (\varphi_8, 0)$ and
$(\varphi_9, 0)
(\varphi_7<\varphi_1^*<\varphi_8<\varphi_2^*<\varphi_9)$.
$(\varphi_7, 0)$ and $(\varphi_9, 0)$ are two center points, while
$(\varphi_8, 0)$ is a saddle point.
\end{theorem}

In this paper, we only consider the case $c>0$ because in the case
$c<0$ we will get analogous result. In order to give the details of
the bifurcation, we fix the parameter $a=-1$. Thus, we obtain the
following two bifurcation curves of system (\ref{eq2.7}).
\begin{equation}
\nonumber L_1:
b=-\frac{2c^3}{27}+\frac{2c^2}{27}\sqrt{c^2-3}-\frac{2}{9}\sqrt{c^2-3}+\frac{c}{3},
\end{equation}
\begin{equation}
\nonumber L_2:
b=-\frac{2c^3}{27}-\frac{2c^2}{27}\sqrt{c^2-3}+\frac{2}{9}\sqrt{c^2-3}+\frac{c}{3}.
\end{equation}

\begin{figure}[h]
\centering \subfloat
{\includegraphics[height=1.8in,width=2.8in]{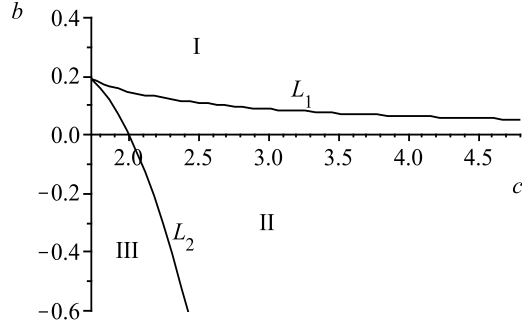}}
 \caption{The
bifurcation sets and bifurcation curves of system (\ref{eq2.7}) for
the parameter $c>\sqrt{3}$.}\label{fig1}
\end{figure}

\begin{figure}[h]
\centering \subfloat[]
{\includegraphics[height=1.5in,width=2.0in]{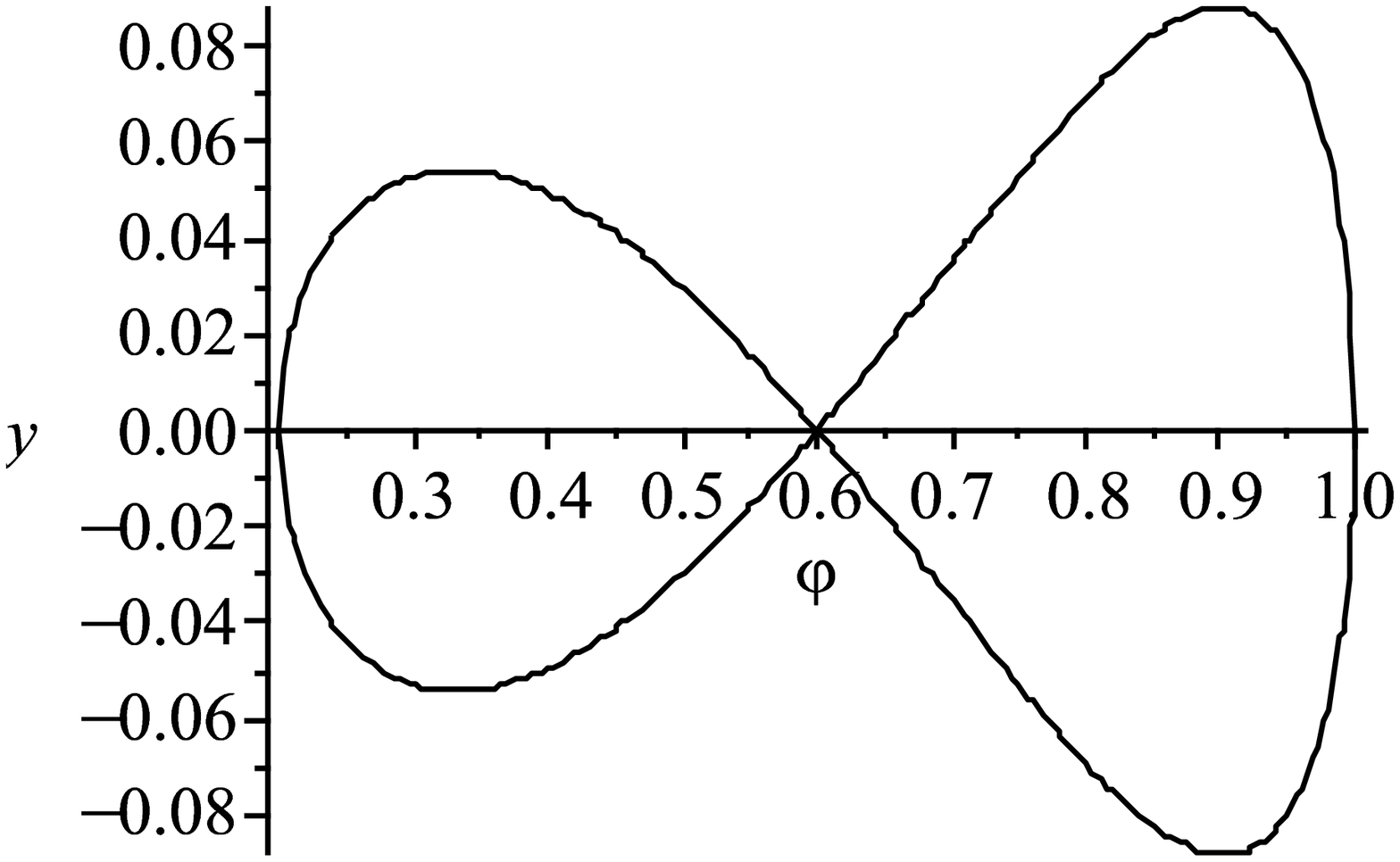}}\hspace{0.1\linewidth}
\subfloat[]{\includegraphics[height=1.5in,width=2.0in]{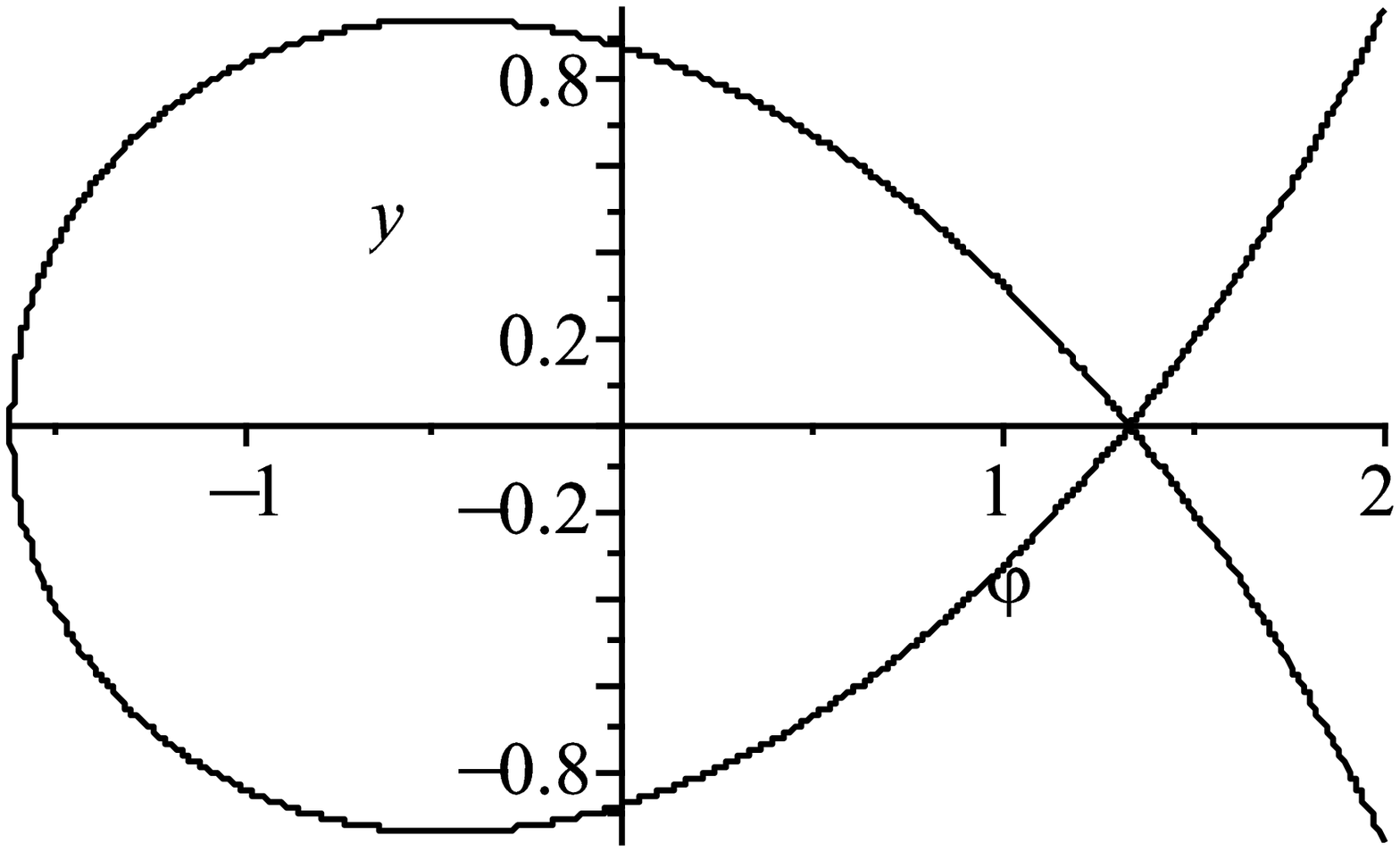}}\\
\caption{The phase portraits of system (\ref{eq2.7}) when the
parameters $c>\sqrt{3}$, $a=-1$ and
$b=-\frac{2c^3}{27}+\frac{c}{3}$. (a) $c=1.8$; (b) $c=4$.
}\label{fig2}
\end{figure}

The above bifurcation curves divide the parameter space into three
regions (see Fig. \ref{fig1}) in which different phase portraits
exist. By theorem \ref{th1}, we can see that only in regions (II),
can system (\ref{eq2.7}) has saddle points. See Fig. \ref{fig2} for
an example of the corresponding phase portraits.


\section{Solitary-wave solutions to system (\ref{eq1.3})}
 \label{}
 \setcounter{equation}{0}

From the discussions in Section 2, we can see that, when the
parameters $a=-1$, $(b, c)\in (\mathrm{II})$, system (\ref{eq2.7})
has infinite many saddle points. So there are infinite many
homoclinic orbits and system (\ref{eq1.3}) has infinite many
solitary-wave solutions accordingly.

In order to obtain exact expressions for solitary-wave solutions, we
fix $b=-\frac{2c^3}{27}+\frac{c}{3}$.

\noindent \textbf{Case I:} $\sqrt{3}<c <3$

In this case, there are two homoclinic orbits connecting with the
saddle point $(\frac{c}{3}, 0)$, see Fig. \ref{fig2}(a) for an
example. The two homoclinic orbits of system (\ref{eq2.7}) or
(\ref{eq2.5}) can be expressed respectively as
\begin{equation}
\label{eq3.1}  y = \pm\frac{( \varphi-\frac{c}{3})\sqrt {-\varphi ^2
+ \frac{2c}{3} \varphi + \frac{5c^2}{9}-2} }{\varphi-c} \quad
\textrm{for} \quad \varphi_-\leq\varphi \leq\frac{c}{3},
\end{equation}
\begin{equation}
\label{eq3.2}  y = \pm\frac{( \varphi-\frac{c}{3})\sqrt {-\varphi ^2
+ \frac{2c}{3} \varphi + \frac{5c^2}{9}-2} }{\varphi-c} \quad
\textrm{for} \quad \frac{c}{3}\leq\varphi \leq \varphi_+,
\end{equation}
where $\varphi_\pm=\frac{1}{3}(c\pm\sqrt{6c^2-18})$.

Substituting (\ref{eq3.1}), (\ref{eq3.2}) into the first equation of
system (\ref{eq2.5}), respectively, and integrating along the
corresponding homoclinic orbit, we have
\begin{equation}
\label{eq3.3} \int_{\varphi_-}^{\varphi} {\frac{s-c}{(
s-\frac{c}{3})\sqrt {-s^2 + \frac{2c}{3} s + \frac{5c^2}{9}-2} } }ds
= \frac{\sqrt{2}}{2}|\xi| ,
\end{equation}
\begin{equation}
\label{eq3.4} \int_\varphi^{\varphi_+}
{\frac{s-c}{(s-\frac{c}{3})\sqrt {-s^2 + \frac{2c}{3} s +
\frac{5c^2}{9}-2} } }ds = -\frac{\sqrt{2}}{2} |\xi|,
\end{equation}
It follows from (\ref{eq3.3}), (\ref{eq3.4}) that
\begin{equation}
\label{eq3.5}\frac{\pi}{2}+
\arctan(\alpha(\varphi))+\frac{2c}{\sqrt{6c^2-18}} \ln
(\beta(\varphi))= \frac{\sqrt{2}}{2}|\xi|, \quad
\varphi_-\leq\varphi \leq\frac{c}{3},
\end{equation}
and
\begin{equation}
\label{eq3.6}
\frac{\pi}{2}-\arctan(\alpha(\varphi))-\frac{2c}{\sqrt{6c^2-18}} \ln
(-\beta(\varphi))= \frac{\sqrt{2}}{2}|\xi|,\quad
\frac{c}{3}\leq\varphi \leq \varphi_+,
\end{equation}
where
\begin{equation}
\label{eq3.7}
\alpha(\varphi)=\frac{3\varphi-c}{\sqrt{-9\varphi^2+6c\varphi+5c^2-18}}$$,
\end{equation}
\begin{equation}
\label{eq3.8}
\beta(\varphi)=\frac{\sqrt{6c^2-18}+\sqrt{-9\varphi^2+6c\varphi+5c^2-18}}{c-3\varphi}.
\end{equation}

Therefore, we obtain two solitary-wave solutions to system
(\ref{eq1.3}) in the following parametric forms:
\begin{equation}
\label{eq3.9} \left\{ {\begin{array}{l}
 \xi=\pm \sqrt{2}(\frac{\pi}{2}+\arctan(\alpha(\varphi))+\frac{2c}{\sqrt{6c^2-18}} \ln
(\beta(\varphi)), \\
 \varphi=\varphi,
\\
 \end{array}} \right. \;\quad (\varphi_-\leq\varphi \leq\frac{c}{3}),
\end{equation}
\begin{equation}
\label{eq3.10} \left\{ {\begin{array}{l}
 \xi=\pm \sqrt{2}(\frac{\pi}{2}+\arctan(\alpha(\varphi))+\frac{2c}{\sqrt{6c^2-18}} \ln
(\beta(\varphi)), \\
 \psi=\frac{g_2}{\varphi-c},
\\
 \end{array}} \right. \;\quad  (\varphi_-\leq\varphi \leq\frac{c}{3}),
\end{equation}
and
\begin{equation}
\label{eq3.11} \left\{ {\begin{array}{l}
 \xi=\pm \sqrt{2}(\frac{\pi}{2}-\arctan(\alpha(\varphi))-\frac{2c}{\sqrt{6c^2-18}} \ln
(-\beta(\varphi)), \\
 \varphi=\varphi,
\\
 \end{array}} \right. \; (\frac{c}{3}\leq\varphi \leq \varphi_+),
\end{equation}
\begin{equation}
\label{eq3.12} \left\{ {\begin{array}{l}
 \xi=\pm \sqrt{2}(\frac{\pi}{2}-\arctan(\alpha(\varphi))-\frac{2c}{\sqrt{6c^2-18}} \ln
(-\beta(\varphi)), \\
 \psi=\frac{g_2}{\varphi-c},
\\
 \end{array}} \right. \; (\frac{c}{3}\leq\varphi \leq \varphi_+).
\end{equation}

\begin{figure}[h]
\centering \subfloat[]
{\includegraphics[height=1.5in,width=2.2in]{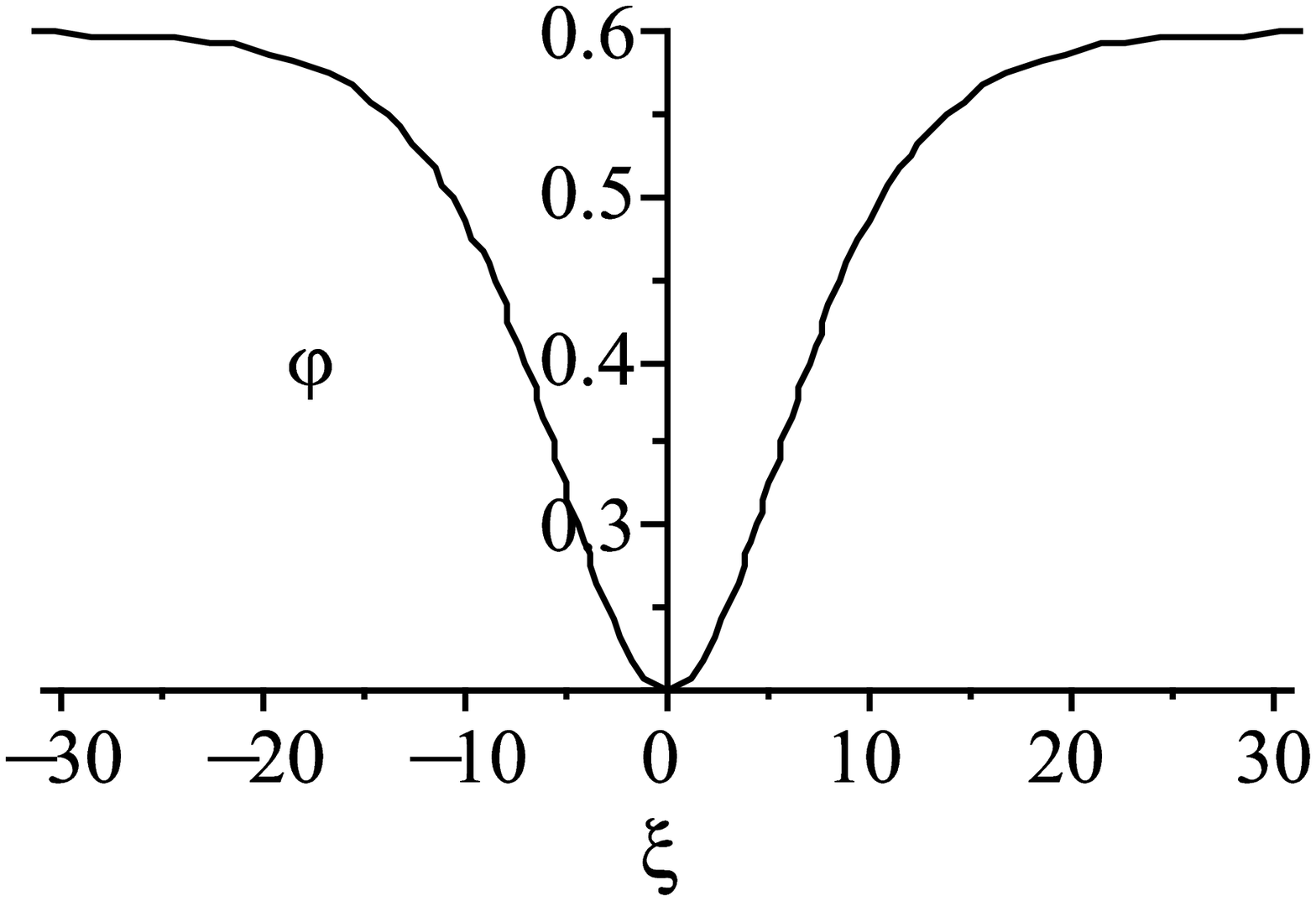}}\hspace{0.15\linewidth}
\subfloat[]{\includegraphics[height=1.5in,width=2.2in]{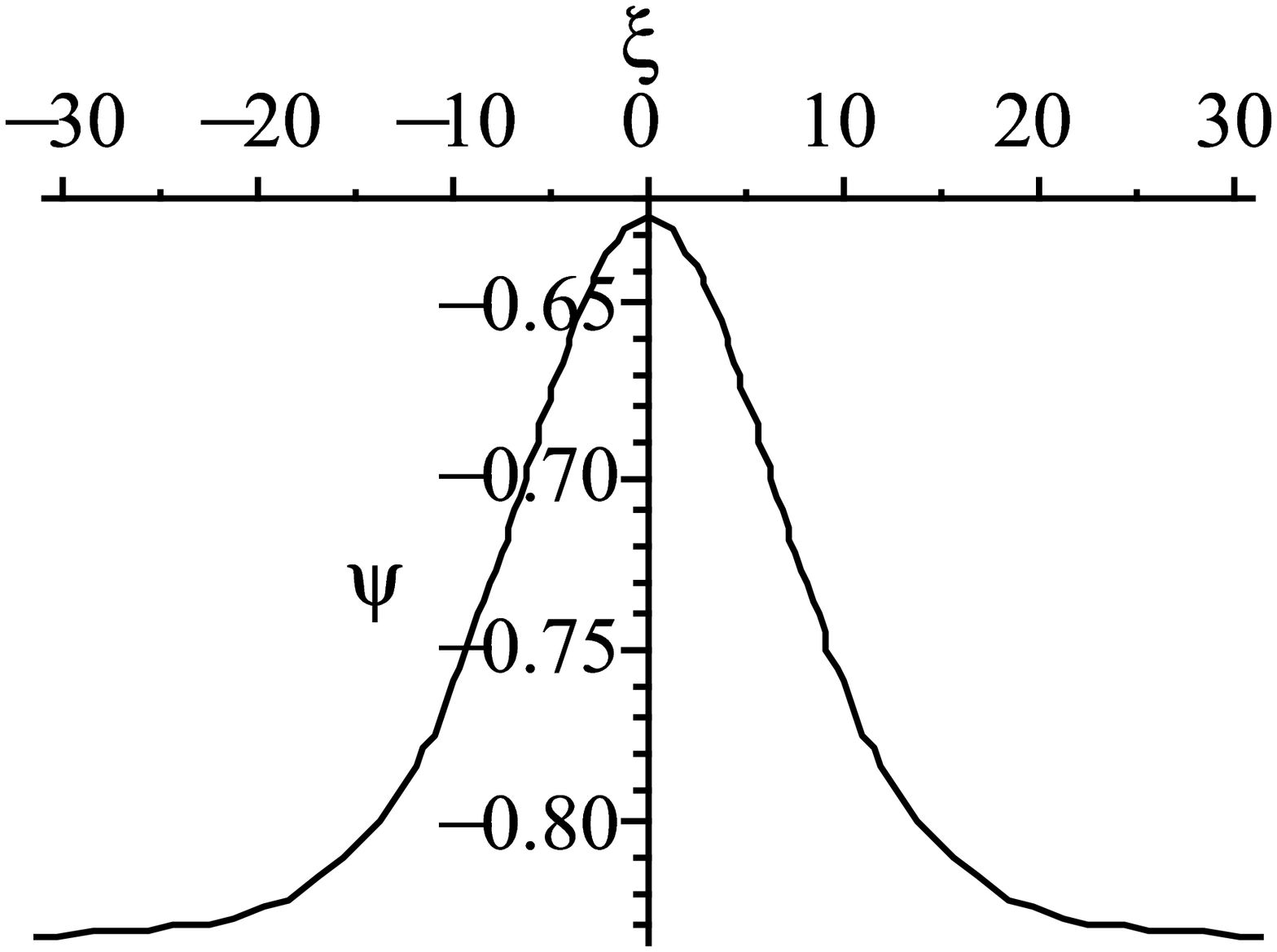}}\\
\subfloat[]
{\includegraphics[height=1.5in,width=2.2in]{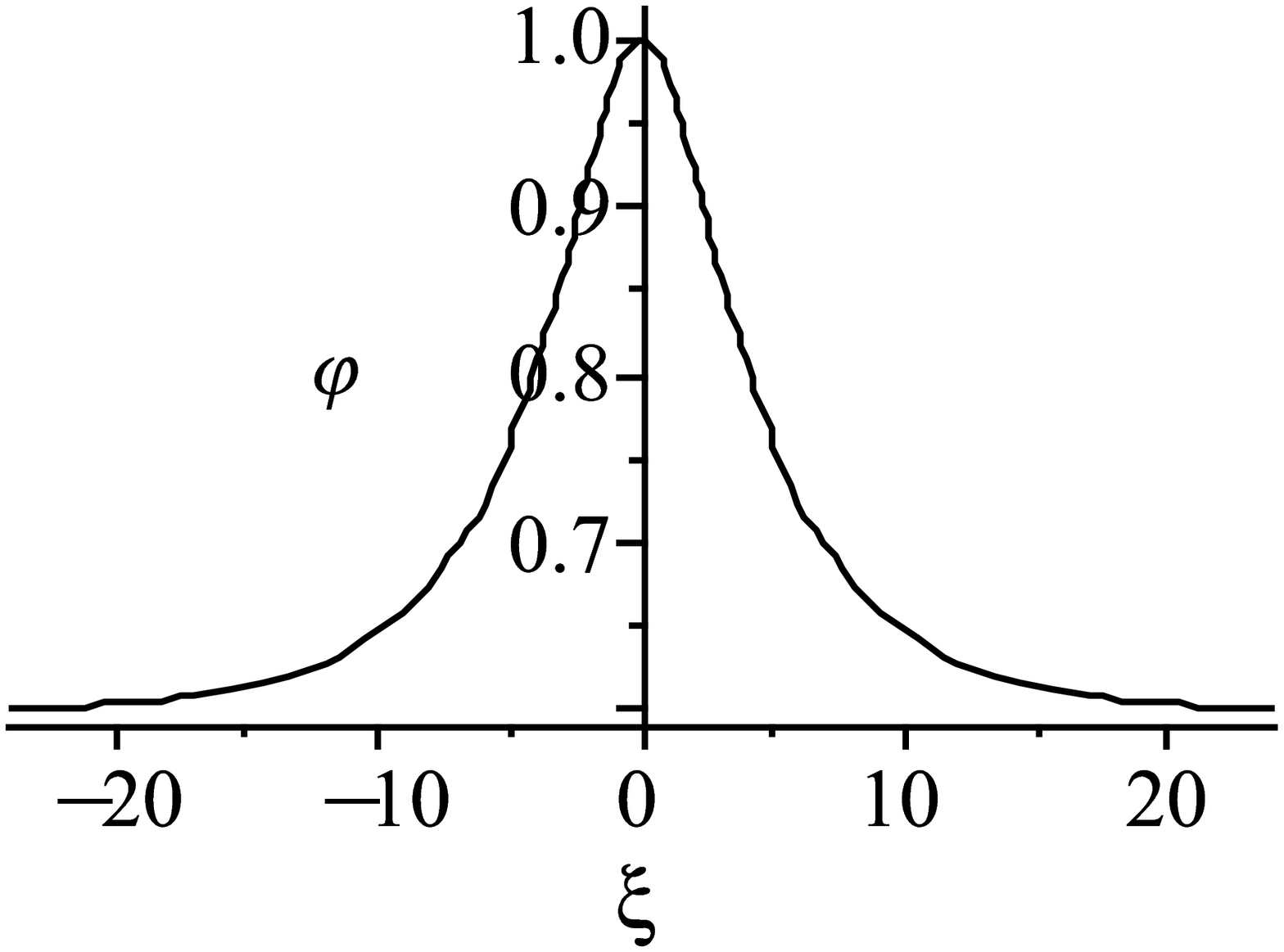}}\hspace{0.15\linewidth}
\subfloat[]{\includegraphics[height=1.5in,width=2.2in]{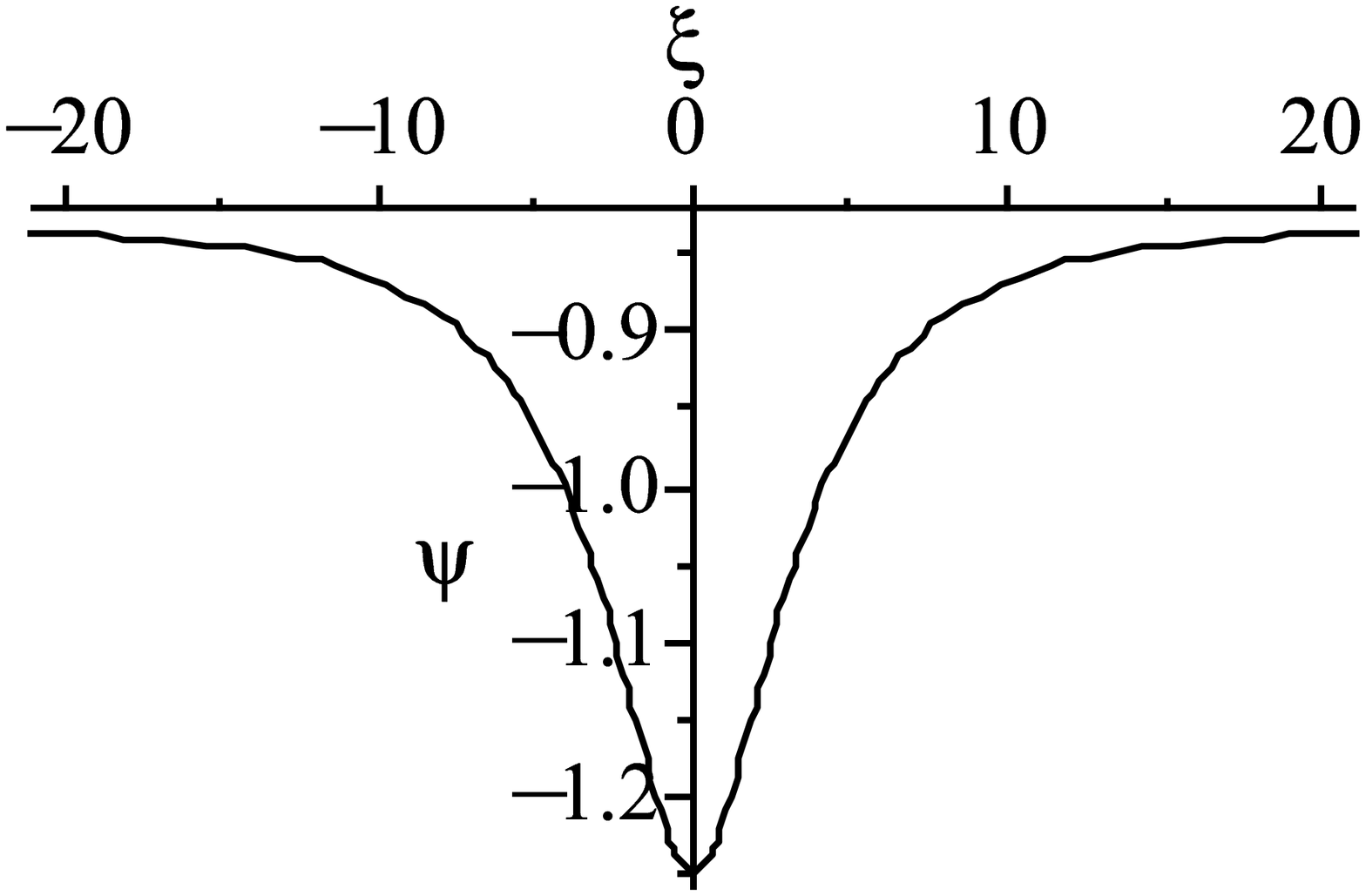}}\\
\caption{Solitary-wave solutions to system (\ref{eq1.3}) when the
parameters $c=1.8$, $a=-1$, $b=0.168000$ and $g_2=1$.}\label{fig3}
\end{figure}

Now we take a set of data and employ Maple to display the graphs of
the above obtained solitary-wave solutions in Fig. \ref{fig3}.

 \noindent \textbf{Case II:} $c\geq 3$

In this case, there is one homoclinic orbit connecting with the
saddle point $(\frac{c}{3}, 0)$, see Fig. \ref{fig2}(b) for an
example.

\begin{figure}[h]
\centering \subfloat[]
{\includegraphics[height=1.5in,width=2.2in]{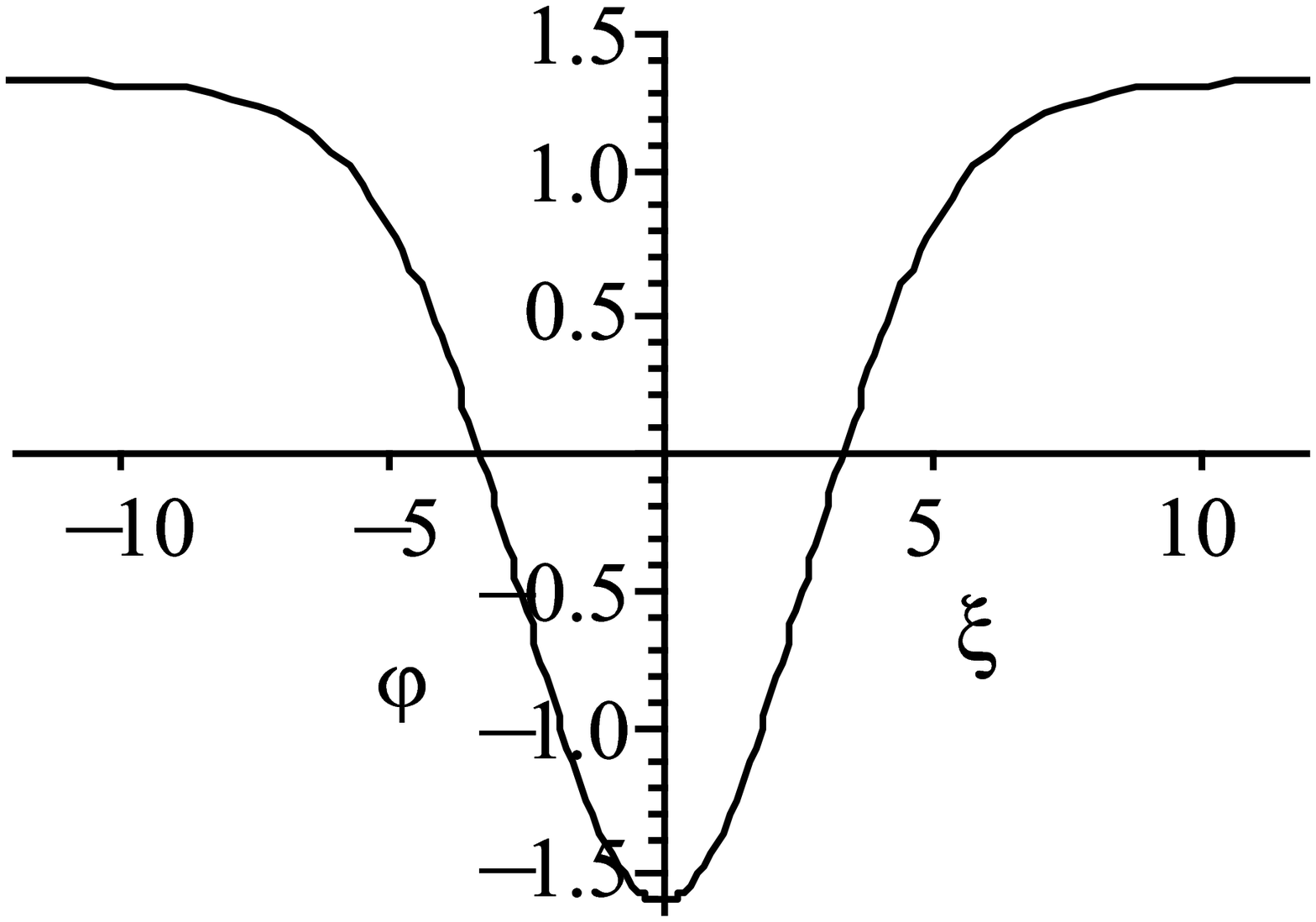}}\hspace{0.1\linewidth}
\subfloat[]{\includegraphics[height=1.5in,width=2.2in]{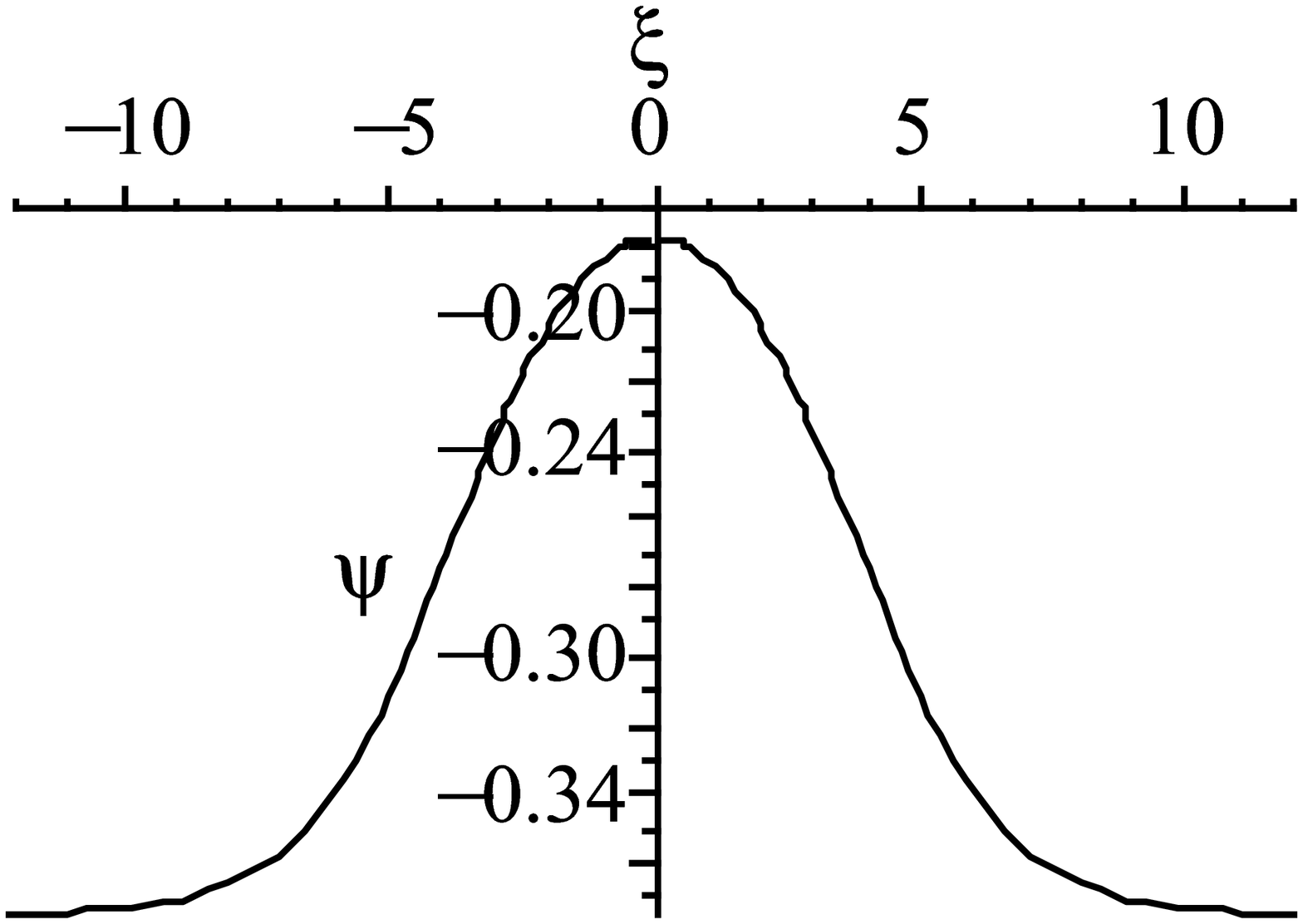}}\\
\caption{Solitary-wave solution to system (\ref{eq1.3}) when the
parameters $c=4$, $a=-1$, $b=-3.407407$ and $g_2=1$.}\label{fig4}
\end{figure}

Similar to the Case I, we can obtain a solitary-wave solution to
system (\ref{eq1.3}), given as (\ref{eq3.9}), (\ref{eq3.10}). A
typical such solution is shown in Fig. \ref{fig4}.

\section{Conclusion}
 \label{}
 \setcounter{equation}{0}
In summary, by using the bifurcation method, we obtain analytic
expressions for smooth solitary-wave solutions to a dual equation of
the Kaup-Boussinesq system (\ref{eq1.3}). The results of this paper
suggest that, in addition to solving many single-component partial
differential equations, the bifurcation method can be used to obtain
travelling-wave solutions of two-component systems.

\end{document}